\documentclass[aps,prd,preprint,floatfix,nofootinbib,groupedaddress,showpacs]
{revtex4}
\usepackage{graphicx}
\usepackage{amsmath}
\usepackage{latexsym}
\usepackage{here}
\usepackage{array}
\usepackage{dcolumn}
\usepackage{bm}

\def\lsim{\mathrel{\rlap{\raise 2.5pt \hbox{$<$}}\lower 2.5pt
\hbox{$\sim$}}}

\begin{document}
\title{Second-class current effects from isospin breaking in $\tau\to\omega\pi\nu_\tau$}

\author{N. Paver}
\affiliation{Department of Physics, University of Trieste, 34100 Trieste, Italy \\
and \\
INFN-Sezione di
Trieste, 34100 Trieste, Italy}

\author{Riazuddin}
\affiliation{National Centre for Physics,\\ Quaid-i-Azam University Campus,\\
Islamabad, Pakistan}

\date{\today}

\begin{abstract}
Second-class weak currents can in the standard model be induced by chiral-symmetry breaking. In the specific case of the decay ${\tau\to\omega\pi\nu_\tau}$, dominated by the first-class vector current with the $\rho$ quantum numbers, such effects would manifest themselves by small axial vector (or, generally, non-vector) contributions to the decay rate. We present an attempt to estimate such effects, based on a vector and axial-vector dominance model of the relevant matrix elements supplemented by $\omega-\rho$ mixing. We also give an indication on the amplitude directly mediated by $b_1(1235)\to\omega\pi$, in principle also allowed in the standard model by isotopic spin violation.
\end{abstract}


\maketitle
The weak currents coupled to $W^\pm$ in semileptonic decays of hadrons composed of $u$ and $d$ quarks can be classified in terms of parity and $G$-parity as follows:  first-class currents with $J^{PG}=0^{--}, 1^{-+}, 1^{+-}$; second-class currents with $J^{PG}=0^{+-}, 1^{++}$  \cite{Weinberg:1958ut,Berger:1987ku}. In the standard model with isospin (hence $G$-parity) conservation only first-class currents exist, and the transition ${\tau\to\omega\pi\nu_\tau}$ would proceed {\it via} a P-wave transition mediated by the $1^{-+}$ `$\rho$-like' vector current. In this situation, the contributions of an S- or a D-wave amplitude would unambiguously signal a $J^{PG}=1^{++}$ second-class, non-standard axial current with the same quantum numbers as the meson $b_1(1235)$ \cite{Leroy:1977pq}. Consequently, ${\tau\to\omega\pi\nu_\tau}$ has been considered as a sensitive test for the existence of second-class currents. With 
${\rm Br}(\tau\to\omega\pi\nu_\tau)=(1.99\pm 0.08)
\times 10^{-2}$ \cite{PDG}, the current experimental upper limit is for this decay: ${\rm Br}({\rm second-class})<1.3\times 10^{-4}$ at 90\% CL \cite{Aubert:2009an}.    
\par 
It should be interesting to assess the size of the `second-class' contributions to this decay generated in the standard model by isospin symmetry breaking, this would be useful to establishing the range in the genuine (non-standard) second-class currents coupling constants still allowed by the above mentioned upper limit for an eventual experimental discovery. The numerical estimates presented in the following will be purely phenomenological, in the sense that our modeling of isospin-breaking-generated second class currents will rely, to the largest possible extent 
within our knowledge, on input values for the needed coupling constants determined experimentally 
and quoted in~\cite{PDG}.  
\par
Following Ref.~\cite{Decker:1992jy}, we separate the hadronic matrix element of the relevant $V-A$ weak current $J^\mu={\bar \psi}_u\gamma^\mu(1-\gamma_5)\psi_d$ into vector and axial vector parts as follows:
\begin{equation}
\label{ffact}
\langle\omega(k,\eta),\pi(p)\vert J^\mu\vert 0\rangle = 
i V(s)\epsilon^{\mu\alpha\beta\gamma}\eta_\alpha k_\beta p_\gamma + A(s)\left(\eta^\mu-\frac{\eta\cdot p}{s} 
\left(k+p\right)^\mu\right).
\end{equation}
Here: $\eta^\mu$ is the $\omega$ polarization vector, $\eta\cdot k=0$; $s=q^2=(k+p)^2$ is the $\omega\pi$ invariant mass squared; and $V(s)$ and $A(s)$ are the (dominant) vector and the (isospin violation suppressed) `second-class' axial-vector form factors, respectively.\footnote{Actually, the most general expansion would require two more form factors, one axial-vector and the other one scalar~\cite{Decker:1992jy}, but we here limit to the ones that according to our estimates are found to be numerically leading.}
\par 
With $s_0=(M_\omega+M_\pi)^2$ and $\lambda(x,y,z)=x^2+y^2+z^2-2(xy+yz+zx)$, the partial decay width can be written as: 
\begin{eqnarray}
\label{decayrate}
\Gamma(\tau\to\omega\pi\nu_\tau) &=&
\frac{G_F^2\vert V_{ud}\vert^2}{1536\pi^3M_\tau^3}
\int_{s_0}^{M_\tau^2}\frac{ds}{s^2}
\lambda^{1/2}(s,M^2_\omega,M^2_\pi) 
\, (M_\tau^2-s)^2\, (M^2_\tau+2s)
\nonumber \\
&\times& \left[\lambda(s,M^2_\omega,M^2_\pi)
\vert V(s)\vert^2+
\frac{\lambda(s,M^2_\omega,M^2_\pi)+12sM_\omega^2}{2sM^2_\omega} \vert A(s)\vert^2\right].
\end{eqnarray}
\par 
The ``forward-backward'' asymmetry, which essentially counts the difference between numbers of events with positive and negative $\cos\theta$, with $\theta$ the $\pi-\tau$ angle in the $\omega\pi$ rest frame, is determined by the interference:
\begin{equation}
\label{FB}
A_{\rm FB}=\frac{1}{\Gamma(\tau\to\omega\pi\nu_\tau)}
\frac{G_F^2\vert V_{ud}\vert^2}{256\pi^3M_\tau^3}
\int_{s_0}^{M_\tau^2}\frac{ds}{s}\lambda(s,M^2_\omega,M^2_\pi)\, (M^2_\tau-s)^2 \, 
{\mathrm{Re}}[A(s)V^*(s)].
\end{equation}
In order to predict the observables (\ref{decayrate}) and (\ref{FB}), explicit expressions for the form factors $V(s)$ and $A(s)$ are needed. 
\par
Theoretical parametrizations for the dominant, first-class, form factor $V(s)$ mostly rely on vector meson exchange, see, for example, Refs.~\cite{Decker:1992jy,Volkov:2012gv}.
We refer to the experimental resonance analysis of ${\tau\to\omega\pi\nu_\tau}$ of  Ref.~\cite{Edwards:1999fj}, and assume the simplified unsubtracted linear combination of 
$\rho\equiv\rho(770) $ and $\rho^\prime\equiv\rho(1450)$ polar forms, see also Ref.~\cite{Datta:2006kd}:  
\begin{equation}
\label{ffV}
V(s)={\displaystyle \frac{\sqrt{2}F_\rho g_{\omega\rho\pi}}{M_\rho^2}\frac{1}{1+\beta_\rho}
\, \left[\frac{M_\rho^2}{M_\rho^2-s} +
\beta_\rho\frac{M_{\rho^\prime}^2}{M_{\rho^\prime}^2-s-
iM_{\rho^\prime}\Gamma_{\rho^\prime}(s)}\right]}.
\end{equation}
In Eq. (\ref{ffV}): $F_\rho\cong M_\rho^2/6$ is the 
$\rho\to e^+e^-$ coupling; for the ${(\omega\rho\pi)}$ 
coupling we take ${g_{\omega\rho\pi}=16.1\, {\rm GeV}^{-1}}$ 
\cite{Edwards:1999fj}; and we choose the value of the constant $\beta_\rho\simeq -0.12$ in order to reproduce, 
from Eq. (\ref{decayrate}), the measured branching ratio of about 2\%. Moreover, the $s$-dependent $\rho^\prime$ width is defined as~\cite{Kuhn:1990ad,Bruch:2004py}:
\begin{equation}
\label{reswidth}
\Gamma(s)=\theta(s-s_0) \frac{M_{\rho^\prime}}{\sqrt s}
\left(\frac{k(s)}{k_{\rho^\prime})}\right)^3
\Gamma_{\rho^\prime}, 
\end{equation}
where $k$ denotes the momentum in the 
$\omega\pi$ c.m. frame. In a sense, Eq.~(\ref{ffV}) resembles the modification 
of the $\rho$ propagator introduced in 
Ref.~\cite{Dumm:2009va}. In Eq.~(\ref{ffV}), the 
width $\Gamma_\rho(s)$ has been omitted, since the  $\rho$-pole is below the threshold $s_0$, but this 
will have little impact on the numerical results. 
Indeed, considering also different, alternative,  parametrizations of the $s$-dependent resonance widths, and the eventual inclusion of the width in the $\rho$ pole, the values of $\beta_\rho$ needed to reproduce the 2\% branching ratio will ultimately range between $-0.12$ and $-0.15$. 
It might be curious to notice that similar values of $\beta_\rho$ have been calculated in vector-dominance 
applications to second-class currents in $\tau$ 
semileptonic decays to $\eta\pi$ and $\eta^\prime\pi$ 
\cite{Paver:2010mz}.\footnote{For simplicity we do not include a non-resonant part of $V(s)$, that for soft pions can be evaluated in chiral perturbation theory~\cite{Davoudiasl:1995ed}.} 
\par 
We model the contribution of the second-class axial 
current to $\tau\to\omega\pi\nu_\tau$ by the transition  of $\tau$ to the axial-vector meson 
$a_1(1260)$, ${\tau\to a_1\nu_\tau}$, followed by  
$a_1\to\rho\pi\to\omega\pi$ {\it via} the isospin 
violating $\rho-\omega$ mixing. Thus, defining 
the $a_1\to\rho\pi$ transition matrix element as  
\begin{equation}
\label{a1decay}
T(a_1(q,\eta)\to\rho(k,\lambda)+\pi(p))=
\left(M_a^2-M_\rho^2\right)\left(\eta\cdot\lambda\right)f_{a\rho\pi}+
2\left(q\cdot\lambda\right)\left(k\cdot\eta\right)
g_{a\rho\pi},
\end{equation}
where $\eta$ and $\lambda$ denote the $a_1$ and $\rho$ polarization vectors, respectively, we would get for the axial form factor $A(s)$ the polar expression  
\begin{equation}
\label{contriba1}
A(s)\vert_{a_1}=\epsilon_{\omega\rho}f_a f_{a\rho\pi}\frac{M_a^2-M_\rho^2}
{M_{a}^2-s-iM_a\Gamma_a(s)}. 
\end{equation}
In Eq. (\ref{contriba1}): $\epsilon_{\omega\rho}$ is the $\omega-\rho$ mixing parameter, and we simply assume 
$\vert\epsilon_{\omega\rho}\vert=3\times 10^{-2}$ from the branching ratio of $\omega\to 2\pi$ - this  also averages, in some cases underestimates, determinations from the timelike 
pion form factor - see, e.g., \cite{Gardner:1997ie}; for the constant $f_a$ defined by $\langle 0\vert{\bar \psi}_u\gamma_\mu\gamma_5 \psi_d\vert a_1(q,\eta)\rangle= 
f_a\eta_\mu$, we take $f_a\simeq 0.2\, {\rm GeV}^2$, assuming the  
${\rm Br}(\tau\to 3\pi\nu_\tau)\simeq 10\%$ 
\cite{PDG} to be saturated by the $a_1$ exchange; 
finally, the values of the constants $f_{a\rho\pi}$ 
and $g_{a\rho\pi}$ can be estimated from the 
$a_1\to\rho\pi$ width. 
\par 
In this regard, the $a_1$ width is rather badly known 
experimentally, $\Gamma_{a_1}$ ranges from 250 to 600  MeV, while the situation is better for the D-wave/S-wave amplitude ratio in the transition $a_1\to\rho\pi$, 
${\rm D/S}=-0.062\pm 0.022$ \cite{PDG}. From this 
ratio, using relations derived in 
Ref.~\cite{Isgur:1988vm}, varying $\Gamma_{a_1}$ in 
the range mentioned above and assuming 
${\rm Br}(a_1\to\rho\pi)$ between 60\% and 100\%, we 
find the values $f_{a\rho\pi}\simeq 3.3 - 5.9$. For 
the coupling constant $g_{a\rho\pi}$ in Eq.~(\ref{a1decay}), that would enter into the second axial form  factor previously alluded to and found numerically suppressed, we would get 
$g_{a\rho\pi}\simeq 0.2 f_{a\rho\pi}$.
\par 
Using Eqs.~(\ref{decayrate}) and (\ref{FB}) with the 
parametrizations (\ref{ffV}) and (\ref{contriba1}) 
and the input parameters varied in the ranges indicated above, we finally obtain the following estimates for the isospin breaking second-class contributions: 
\begin{equation}
\label{a1}
{\rm Br}(\tau\to\omega\pi\nu_\tau)\vert_{a_{1}} 
\simeq (1.6-2.1)\times 10^{-5};\, \, \, \, \, 
\vert A_{\rm FB}\vert\simeq (2.4-4.8)\times 
10^{-3}. 
\end{equation} 
As one can see, the uncertainty is rather large, 
and is mainly due to the extended range where   
the $a_1$ parameters can vary. However, the upper values in Eq.~(\ref{a1}) are the most important  
ones for our purposes, in that they represent estimated limits for eventually observed second-class effects in 
${\tau\to\omega\pi\nu_\tau}$ to be unambiguously considered as genuine, non-standard, signals rather than manifestations of symmetry breaking in the standard model. 
\par 
An additional "second-class" axial-vector contribution from isotopic spin violation can be represented by the $b_1(1235)$ exchange, which we wish to parametrize analogously to Eq.~(\ref{contriba1}). To this purpose, we recall  that gluon corrections to the ``bare'' 
${{\bar u}dW}$ vertex may generate a pseudotensor, divergenceless, coupling proportional to 
$\Delta m=m_d-m_u$, of the form 
\cite{Halprin:1976rs,Gavela:1980hh}  
\begin{equation}
\label{pseudotensor}
{\bar A}_\mu^{II}(x)={\bar g}_{\rm T}\, \partial^\nu\, {\bar\psi}_u(x)
\sigma_{\mu\nu}\gamma_5\psi_d(x) 
\equiv {\bar g}_{\rm T}A_\mu^{II};\, \, \, \, \, \, \,  
{\bar g}_{\rm T}=-\frac{4\alpha_s}{3\pi m}\, \frac{\Delta m}{2m}, 
\end{equation} 
where $m$ is the average quark mass. If in (\ref{pseudotensor}) one literally used current quark masses of the MeV order, the size of ${\bar g}_{\rm T}$ would be very large, of order 5 or more in ${\rm GeV}^{-1}$ units. However, this would be unjustified, because Eq.~(\ref{pseudotensor}) strictly refers to free quarks. As discussed in 
Refs.~\cite{Halprin:1976rs,Gavela:1980hh,Fajfer:1983iu}, 
one expects that for confined quarks the loop integration over the gluon frequencies needed to 
derive this equation cannot run up to infinity, but must be cut-off at a scale appropriate to the hadronic scale. This will decrease the size of ${\bar g}_{\rm T}$ appreciably, in particular down to an  order of magnitude compatible with phenomenological limits on  second-class currents from nuclear $\beta$-decay, see, as an example, Refs.  \cite{Wilkinson:2000gx,Kubodera:1977ai,Minamisono:2011zz}. Accordingly, as a criterion to account for confinement effects in Eq.~(\ref{pseudotensor}), we choose to input there the constituent quark masses 
$M_u\simeq M_d = 350\, {\rm MeV}$, 
$\Delta M=2\, {\rm MeV}$, and $\alpha_s=0.5$. This gives the indicative estimate 
${\bar g}_{\rm T}=-1.7\times 10^{-3}\, {\rm GeV}^{-1}$.
\par
We now need the pseudotensor constant 
$\langle 0\vert{\bar\psi}_u\sigma_{\mu\nu}\gamma_5\psi_d 
\vert b_1(q,\eta)\rangle=if_b(\eta_\mu q_\nu-\eta_\nu q_\mu)$, 
for which we assume the quark-model value $f_b=\sqrt{2}f_a/M_b$ \cite{Gamberg:2001qc}. After contraction with $q^\nu$ as required by the expression 
(\ref{pseudotensor}), with $q^2=M_b^2$ and 
$\eta\cdot q=0$, we obtain for the second-class axial 
current matrix element: 
$\langle 0\vert{\bar A}_\mu^{II}\vert b_1(q,\eta)\rangle = 
{\bar g}_{\rm T}\, f_a\sqrt{2}\, M_b\, \eta_\mu$. With the 
$b_1\to\omega\pi$ matrix element defined similar to 
Eq.~(\ref{a1decay}),     
\begin{equation}
\label{b1decay}
T(b_1(q,\eta)\to\omega(k,\lambda)+\pi(p))=
\left(M_b^2-M_\omega^2\right)\left(\eta\cdot\lambda\right)f_{b\rho\pi}+
2\left(q\cdot\lambda\right)\left(k\cdot\eta\right)
g_{b\rho\pi},
\end{equation}
we finally arrive at the following parametrization for the ``direct'' $b_1$ contribution to the form factor 
$A(s)$:
\begin{equation}
\label{contribb1}
A(s)\vert_{b_1}={\bar g}_{\rm T} M_b \sqrt{2}\, f_a\,  f_{b\omega\pi}\, \frac{M_b^2-M_\omega^2}
{M_{b}^2-s-iM_b\Gamma_b(s)}. 
\end{equation}
With $\Gamma_{b_1}=142\, {\rm MeV}$, dominated by 
the $b_1$-decay into $\omega\pi$, and the D/S amplitude 
ratio 0.277~\cite{PDG}, by a procedure similar to the case of the $a_1$ we obtain the value 
$f_{b\omega\pi}\simeq 5.0\, {\rm GeV}^{-1}$ and in this way we complete the list of inputs needed in 
to numerically exploit Eq.~(\ref{contribb1}). 
Finally, we represent the axial form factor $A(s)$ by the combination of $a_1$ and $b_1$ poles:
\begin{equation}
\label{a1plusb1}
A(s)=A(s)\vert_{a_1}+A(s)\vert_{b_1}.
\end{equation}
\par
One can notice that, according to 
the above numerical estimates, 
the factor multiplying the $b_1$ Breit-Wigner form 
in (\ref{contribb1}) is suppressed with respect to the analogous factor multiplying the $a_1$ pole 
in (\ref{contriba1}) by about $10^{-1}$. Possibly,  this might be an overestimate of the $b_1$ contribution, did we choose for the mass scale in  Eq.~(\ref{pseudotensor}) the hadron mass, for example $m_{b_1}$, instead of the constituent quark masses, a smaller value of ${\bar g}_T$ would have followed. Using Eq.~(\ref{a1plusb1}), and the input values obtained above, we would find for the isospin-breaking induced second-class effects:
\begin{equation}
\label{final}
{\rm Br}(\tau\to\omega\pi\nu_\tau)\vert_{A(s)}\simeq 
(2.3-2.8)\times 10^{-5}, \, \, \, \, \, \, 
\vert A_{\rm FB}\vert\simeq (2.6-5.3)\times 10^{-3},
\end{equation}  
to be compared to the current upper limit on the 
second-class branching ratio of the order of $10^{-4}$ mentioned at the beginning. The numbers in (\ref{final}) fall well-below that limit, and indicate that there still is ample room for 
an eventual discovery of second-class 
currents in the decay $\tau\to\omega\pi\nu_\tau$, before 
standard model, isospin breaking, effects are met. As 
regards the current situation with the pseudotensor genuine second-class current $A_\mu^{II}$ defined in (\ref{pseudotensor}), in $\beta$-decay it contributes 
as: $\langle p\vert A_\mu^{II}\vert n\rangle= 
g_{\rm T}^{(\beta)}
{\bar u}(p)i\sigma_{\mu\nu}\gamma_5 q^\nu u(n)$. Barring cancellations with other, non-pseudotensor, forms of second-class currents, the limits from different observables could be summarized by 
$\vert g_{\rm T}^{(\beta)}\vert\leq (2-5)\times 10^{-1}\, 
{\rm GeV}^{-1}$ (see \cite{Minamisono:2011zz} and references therein). The nucleon matrix element of $A_\mu^{II}$ is 
suppressed by the smallness of the four-momentum $q$ in $\beta$-decay. In $\tau\to\omega\pi\nu_\tau$ the momentum $q$ 
is not small, this might enhance the sensitivity of this decay to $A_\mu^{II}$. In fact, defining for a comparison a 
scale $g_{\rm T}^{(\omega\pi)}$ analogous to 
$g_{\rm T}^{(\beta)}$ and with same dimensions, and introducing it in 
Eq. (\ref{contribb1}) in place of ${\bar g}_{\rm T}$, 
the current upper limit on the axial-vector branching ratio 
quoted at the beginning can exclude genuine 
second-class currents at the level 
$\vert g_{\rm T}^{(\omega\pi)}\vert\leq 2\times 10^{-2}\,{\rm GeV}^{-1}$.

\goodbreak
\leftline{\bf Acknowledgments}
\par\noindent
This research has been partially supported by funds of the University of Trieste. One of the
authors (R) would like to thank the Abdus Salam ICTP
for hospitality. Our thanks to Ali Paracha for help 
in the numerical work. 

\goodbreak


\begin{thebibliography}{99}

\bibitem{Weinberg:1958ut}
S.~Weinberg,
Phys.\ Rev.\ {\bf 112}, 1375 (1958).
  
\bibitem{Berger:1987ku}
E.~L.~Berger and H.~J.~Lipkin,
Phys.\ Rev.\ Lett.\  {\bf 59}, 1394 (1987).

\bibitem{Leroy:1977pq}
C.~Leroy and J.~Pestieau,
Phys.\ Lett.\  B {\bf 72}, 398 (1978); 
R.~Tegen,
Z.\ Phys.\  C {\bf 7}, 121 (1981).

\bibitem{PDG}
K.~Nakamura {\it al.} [Particle Data Group], 
J.\ Phys.\ G\ {\bf 37}, 075021 (2010).   

\bibitem{Aubert:2009an} 
  B.~Aubert {\it et al.}  [BABAR Collaboration],
Phys.\ Rev.\ Lett.\  {\bf 103}, 041802 (2009)  [arXiv:0904.3080 [hep-ex]].  

\bibitem{Decker:1992jy} 
R.~Decker and E.~Mirkes,
Z.\ Phys.\ C {\bf 57}, 495 (1993).  

\bibitem{Volkov:2012gv} 
M.~K.~Volkov, A.~B.~Arbuzov and D.~G.~Kostunin, 
arXiv:1204.4537 [hep-ph];   
Z.~-H.~Guo, 
Phys.\ Rev.\ D {\bf 78}, 033004 (2008)  [arXiv:0806.4322 [hep-ph]];   
A.~Flores-Tlalpa and G.~L\'opez-Castro, 
Phys.\ Rev.\ D {\bf 77}, 113011 (2008)  [arXiv:0709.4039 [hep-ph]];   
S.~Fajfer, K.~Suruliz and R.~J.~Oakes,
Phys.\ Rev.\ D {\bf 46}, 1195 (1992).  

\bibitem{Edwards:1999fj} 
K.~W.~Edwards {\it et al.} [CLEO Collaboration],
Phys.\ Rev.\ D {\bf 61}, 072003 (2000)  [hep-ex/9908024].  

\bibitem{Datta:2006kd} 
A.~Datta, K.~Kiers, D.~London, P.~J.~O'Donnell and A.~Szynkman,
Phys.\ Rev.\ D {\bf 75}, 074007 (2007)  [Erratum-ibid.\ D {\bf 76}, 079902 (2007)]  [hep-ph/0610162]. 

\bibitem{Kuhn:1990ad}
J.~H.~K\"uhn and A.~Santamaria,
Z.\ Phys.\ C {\bf 48}, 445 (1990).

\bibitem{Bruch:2004py}
C.~Bruch, A.~Khodjamirian and J.~H.~Kuhn,
Eur.\ Phys.\ J.\ C {\bf 39}, 41 (2005) 
[arXiv:hep-ph/0409080].

\bibitem{Dumm:2009va}
D.~G.~Dumm, P.~Roig, A.~Pich and J.~Portoles,
Phys.\ Lett.\ B {\bf 685}, 158 (2010) 
[arXiv:0911.4436 [hep-ph]].

\bibitem{Paver:2010mz} 
N.~Paver and Riazuddin,
Phys.\ Rev.\ D {\bf 82}, 057301 (2010) [arXiv:1005.4001 [hep-ph]];  
Phys.\ Rev.\ D {\bf 84}, 017302 (2011)  [arXiv:1105.3595 [hep-ph]].  

\bibitem{Davoudiasl:1995ed} 
H.~Davoudiasl and M.~B.~Wise,
Phys.\ Rev.\ D {\bf 53}, 2523 (1996)  [hep-ph/9509414].  

\bibitem{Gardner:1997ie} 
S.~Gardner and H.~B.~O'Connell,
Phys.\ Rev.\ D {\bf 57}, 2716 (1998)  [Erratum-ibid.\ D {\bf 62}, 019903 (2000)]  [hep-ph/9707385];   
C.~E.~Wolfe and K.~Maltman,
Phys.\ Rev.\ D {\bf 83}, 077301 (2011)  [arXiv:1011.4511 [hep-ph]]. 

\bibitem{Isgur:1988vm} 
N.~Isgur, C.~Morningstar and C.~Reader,
Phys.\ Rev.\ D {\bf 39}, 1357 (1989).  

\bibitem{Halprin:1976rs} 
A.~Halprin, B.~W.~Lee and P.~Sorba,
Phys.\ Rev.\ D {\bf 14}, 2343 (1976).  

\bibitem{Gavela:1980hh} 
M.~B.~Gavela, A.~Le Yaouanc, L.~Oliver, O.~Pene and J.~C.~Raynal,
Phys.\ Rev.\ D {\bf 22}, 2906 (1980).  

\bibitem{Fajfer:1983iu} 
S.~Fajfer and R.~J.~Oakes,
Phys.\ Rev.\ D {\bf 28}, 2881 (1983).  

\bibitem{Wilkinson:2000gx} 
D.~H.~Wilkinson,
Eur.\ Phys.\ J.\ A {\bf 7}, 307 (2000).  

\bibitem{Kubodera:1977ai} 
K.~Kubodera, J.~Delorme and M.~Rho,
Phys.\ Rev.\ Lett.\  {\bf 38}, 321 (1977).  

\bibitem{Minamisono:2011zz} 
K.~Minamisono, T.~Nagatomo, K.~Matsuta, C.~D.~P.~Levy, Y.~Tagishi, M.~Ogura, M.~Yamaguchi and H.~Ota {\it et al.},
Phys.\ Rev.\ C {\bf 84}, 055501 (2011).  

\bibitem{Gamberg:2001qc} 
L.~P.~Gamberg and G.~R.~Goldstein,
Phys.\ Rev.\ Lett.\ {\bf 87}, 242001 (2001)  [hep-ph/0107176].  

\end{thebibliography}
\end{document}